\documentclass[letterpaper, 10 pt, conference]{ieeeconf}
\IEEEoverridecommandlockouts
\usepackage{cite}
\usepackage{amsmath,amssymb,amsfonts}
\usepackage{algorithmic}
\usepackage{graphicx}
\usepackage{svg}       
\usepackage{textcomp}
\usepackage{xcolor}

\usepackage{pstricks}
\usepackage{tikz}

\def\BibTeX{{\rm B\kern-.05em{\sc i\kern-.025em b}\kern-.08em
    T\kern-.1667em\lower.7ex\hbox{E}\kern-.125emX}}

\begin{document}

\title{
Improved Voltage Regulation with Optimal Design of Decentralized Volt-VAr Control
\thanks{This material is based upon work supported by the U.S. Department of Energy’s Office of Energy Efficiency and Renewable Energy (EERE) under the Solar Energy Technologies Office Award Number DE-EE0010147 and the Community Energyshed Design initiative, award number DE-EE0010407. 
The views expressed herein do not necessarily represent the views of the U.S. Department of Energy or the United States Government. 
The authors also thank Cyril Brunner of VEC and Dan Kopin of VELCO for their data sharing and continued collaboration.}
}

\author{\IEEEauthorblockN{Dan Russell,
Dakota Hamilton, Mads Almassalkhi and
Hamid R. Ossareh}
\IEEEauthorblockA{Department of Electrical \& Biomedical Engineering,
University of Vermont\\
Burlington VT, United States\\
Email: daniel.russell@uvm.edu,
dakota.hamilton@uvm.edu,
malmassa@uvm.edu,
hamid.ossareh@uvm.edu}}

\title{\LARGE \bf
Improved Voltage Regulation with Optimal Design \\ of Decentralized Volt-VAr Control
}

\author{Daniel Russell, Dakota Hamilton, Mads R. Almassalkhi, and Hamid R. Ossareh$^{1}$
\thanks{This material is based upon work supported by the U.S. Department of Energy’s Office of Energy Efficiency and Renewable Energy (EERE) under the Solar Energy Technologies Office Award Number DE-EE0010147 and the Community Energyshed Design initiative, award number DE-EE0010407. 
The views expressed herein do not necessarily represent the views of the U.S. Department of Energy or the United States Government. 
The authors also thank Cyril Brunner of VEC and Dan Kopin of VELCO for their data sharing and continued collaboration.}
\thanks{$^{1}$ The authors are with the Department of Electrical and Biomedical Engineering, University of Vermont, Burlington, VT, USA
{\tt\small \{djrussel, dhamilt6, malmassa, hossareh\}@uvm.edu}}%
}

\maketitle

\begin{abstract}
Integration of distributed energy resources has created a need for autonomous, dynamic voltage regulation. 
Decentralized Volt-VAr Control (VVC) of grid-connected inverters presents a unique opportunity for voltage management but, if designed poorly, can lead to unstable behavior when in feedback with the grid. 
We model the grid-VVC closed-loop dynamics with a linearized power flow approach, leveraging historical data, which shows improvement over the commonly used LinDistFlow model. 
This model is used to design VVC slopes by minimizing steady-state voltage deviation from the nominal value, subject to a non-convex spectral radius stability constraint, which has not been previously implemented within this context. 
We compare this constraint to existing convex restrictions and demonstrate, through simulations on a realistic feeder, that  using the spectral radius results in more effective voltage regulation.
\end{abstract}


\section{Introduction}

\begin{figure}[t]
\centerline{\includegraphics[width=0.5\textwidth]{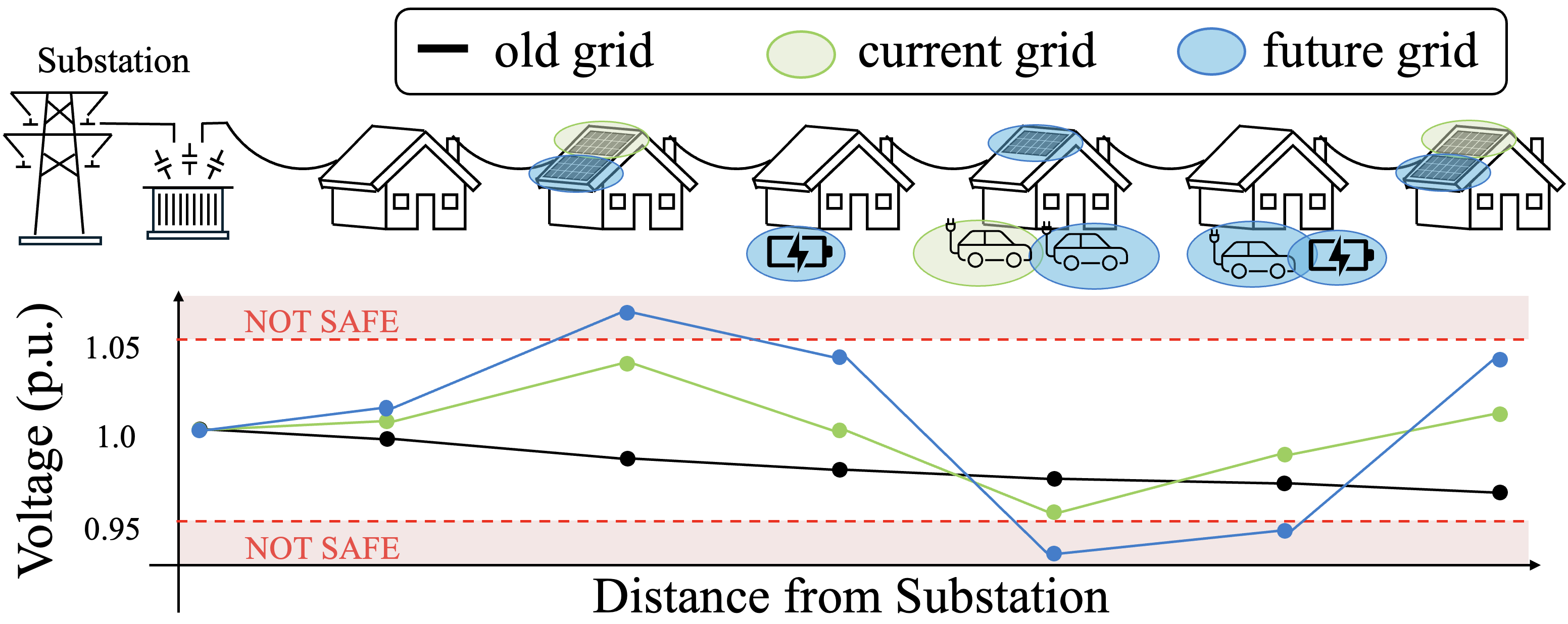}}
\caption{As proliferation of DERs increase, voltage profiles on distribution networks change faster and by larger amounts. Note that this is a cartoon for illustrative purposes only, not real data. }
\label{fig:Motivation}
\end{figure}

The decreasing cost of solar photovoltaics (PV) is leading to rapid deployment of this technology in the electric distribution grid~\cite{ritchie2023,ressolar2025}. 
However, in regions where high solar PV adoption is combined with new loads like electric vehicles and other distributed energy resources~(DERs), fast changes in generation and demand can cause distribution system voltages to change rapidly and significantly (as illustrated in Fig.~\ref{fig:Motivation}).
This variability in distribution grid voltages can negatively impact grid reliability~\cite{tan2007}. 
Fortunately, inverter-interfaced DERs can assist in managing system voltages by injecting or absorbing reactive power at their point-of-connection with the grid, known as Volt-VAr Control~(VVC)~\cite{grainger1985}.
The objective of this paper is to present a framework for designing optimal VVC rules that not only regulate voltages effectively but also guarantee dynamic stability when interconnected with the grid.

Implementations of VVC fit into three categories: \textit{centralized}, \textit{distributed}, and \textit{decentralized} control. 
\textit{Centralized} VVC is often capable of providing globally optimal results (e.g. minimum voltage deviation or minimum power losses) because of its presumed knowledge of the grid at all times~\cite{gush}. 
However, as pointed out in~\cite{worthmann}, access to data at that scale is impractical and coordinating all devices directly exposes the grid to cybersecurity threats~\cite{DoECybersecurity}.
To reduce the need for extensive communication networks and central computations, \textit{distributed} control solutions as part of a laminar architecture can rely on less data and/or less frequent communication. 
Strategies like multi-agent reinforcement learning~\cite{liu&wu,sun} and centralized optimization with partial grid information~\cite{gupta} have been applied. 
All of these schemes still require some communication infrastructure, making them harder to implement.
Due to these challenges, we focus on \textit{decentralized} control, which uses devices in a network that only require local information to make control decisions.

\begin{figure}[t]
\centerline{\includegraphics[width=0.5\textwidth]{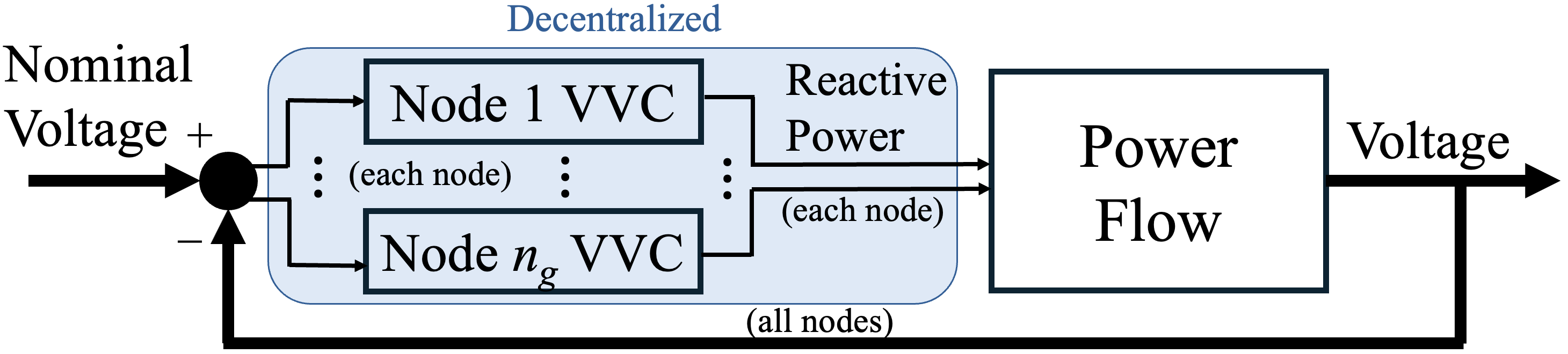}}
\caption{A block diagram showing the interaction between Volt-VAr Control and the grid physics (plant model). Bold lines indicate a vector of measurements at all nodes.}
\label{fig:VVC_BD}
\end{figure}

The IEEE 1547-2018 Standard~\cite{IEEE1547} specifies various decentralized voltage-control policies for grid-connected inverters. Among these, VVC is favored in this study (and many others) as it preserves inverter life-time~\cite{mohamed}. 
The IEEE Standard specifies VVC as a curve, which proportionally relates voltage at a particular node to a reactive power response. 
The interaction between VVC and the grid physics is shown in Fig.~\ref{fig:VVC_BD} and the control curve for this policy can be seen in Fig.~\ref{fig:IEEE-VVC}.
There has been significant study of VVC policies that do not adhere to this curve structure, such as with artificial neural networks~\cite{liNeuralNets} and chance-constrained optimal power flow plus machine learning~\cite{eggli2021}. 
Although these strategies can perform well, lack of standard compliance limits implementability.

Even with an IEEE-compliant control curve, the system in Fig.~\ref{fig:VVC_BD} can exhibit unstable voltages: 
poor VVC design (i.e., steep slopes) and no filtering can cause voltage oscillation or even voltage collapse.
This phenomenon has been analyzed in~\cite{jahangiri}.
While that work does not address optimal design, subsequent studies by~\cite{zhou2021} and~\cite{murzakhanov2024} do consider design, albeit using conservative restrictions of the stability constraint defined by spectral radius. 
In this paper, we address the gap by using the non-convex spectral radius constraint directly.

Additionally, much of the literature proposes an incremental or ``filtered'' approach to VVC, where a low-pass filter is applied at the inverter output. According to~\cite{eggli2021}, this approach results in guaranteed asymptotic stability of a unique equilibrium without needing  restrictive or non-convex stability constraints. 
However, because the IEEE standard does not strictly specify an incremental approach, we choose to study the worst-case scenario with no filtering, as was done in~\cite{wei2023}.

Much of VVC modeling leverages the linear  LinDistFlow~(LDF) model, originally derived from~\cite{BaranWu89} and used in~\cite{zhou2021,feng2024,wei2023,eggli2021,murzakhanov2024} and many more. 
This model makes two key assumptions: first that the loss term in the power flow equations is negligible, and second that voltage magnitude everywhere stays close to 1 per unit~(p.u.). 
Various alternatives to this model have been proposed, including~\cite{dhople2015} which shows an analytical method for approximating power flow as well as data-driven approaches discussed in~\cite{jia2023}. 
We move away from LDF and develop a model derived from a data-based linearization of the power flow that is more accurate within the presented VVC context. 

Accordingly, the contributions of this paper are as follows:
\begin{itemize}
    \item A linearized power flow~(LPF) model using a single historical average operating point is developed for offline design that shows improved accuracy of voltages over the LinDistFlow model.
    \item A novel non-incremental VVC policy design framework is presented as a non-convex optimization problem that accounts for both steady-state voltage deviations and grid-VVC stability, resulting in less conservative controllers.
    \item Simulation-based analysis that validates VVC design on different feeders, including a realistic feeder in Vermont, U.S.A. and compares the approach against prior work. 
\end{itemize} 

The remainder of this paper is laid out as follows. 
Section~\ref{sec:SysModel} describes mathematical modeling of the system in Fig.~\ref{fig:VVC_BD} while Sec.~\ref{sec:stab_crit} describes various stability conditions for this system. 
The optimization problem for VVC design is formulated in Sec.~\ref{sec:opt_design}. 
We describe and validate on a test network and dataset in Sec.~\ref{sec:Network}. 
Performance of the design framework herein is shown in Sec.~\ref{sec:Results} with conclusions drawn in Sec.~\ref{sec:conclusion}.

\section{System Modeling}
\label{sec:SysModel}

For optimal VVC design, we require models of both the grid physics and the VVC. 
We employ a power flow model of the grid that is linear with respect to power injections, but does not neglect losses, as is the case with the oft used LDF.

\subsection{Linearized Power Flow Model} 
\label{subs:LPF}

Consider a single-phase (or single-phase equivalent), radial power distribution network with $n$ nodes (excluding the substation) and generator nodes $\mathcal{N}_\text{g} \subseteq \{1,\hdots, n\}$ with $|\mathcal{N}_\text{g}|=: n_\text{g} \le n$.
We assume the substation voltage magnitude and angle are fixed.
The nonlinear AC power flow can be represented by 
\begin{equation}
\label{eq:ac_pf}
    \mathbf{v} = f_\text{pf}(\mathbf{p}, \mathbf{q})\,, \quad 
    f_\text{pf} : \mathbb{R}^{n}\times\mathbb{R}^{n}  \to \mathbb{R}^n\,,
\end{equation}
where $\mathbf{v}\in\mathbb{R}^n$ denotes the vector of nodal voltage magnitudes, and $\mathbf{p} = \mathbf{p}_\text{g} - \mathbf{p}_\text{d}$ 
and $\mathbf{q} = \mathbf{q}_\text{g} - \mathbf{q}_\text{d}$ 
are vectors of net (generation minus demand) active and reactive power at all nodes, respectively. 
Because demand and generation vary over time, the vector quantities $\mathbf{v}, \mathbf{p}, \mathbf{q}$ are all implicitly functions of time. However, since the power flow model is quasi-steady-state, the explicit dependence on time, $t$, is not shown for clarity of presentation.
The function $f_\text{pf}$ is the mapping between power injections and voltage magnitudes at all nodes, which is equivalent to the \textit{DistFlow} formulation in~\cite{BaranWu89} when the network is radial with loads as P-Q buses.\footnote{P-Q bus refers to a node modeled with a constant power (P and Q) and variable voltage magnitude and angle. Therefore, in a network with all P-Q buses, the voltage can be a function of only P and Q.}

Although voltage phase angles are also an output of the power flow solution, we choose to omit this from $f_\text{pf}$ since we are primarily interested in the control of voltage magnitudes.  

Using a first-order Taylor Series expansion, we linearize~\eqref{eq:ac_pf} around an operating point $(\mathbf{p}_0, \mathbf{q}_0)$.
That is,
\begin{equation}
\label{eq:LPF}
    \mathbf{v} \approx f_\text{pf}(\mathbf{p}_0, \mathbf{q}_0) + \mathbf{J_p}(\mathbf{p} - \mathbf{p}_0) + \mathbf{J_q}(\mathbf{q} - \mathbf{q}_0)\,,
\end{equation}
where Jacobian matrices are derived from
\begin{equation}
\mathbf{J_p} \triangleq \frac{\partial f_\text{pf}}{\partial \mathbf{p}}\Bigg|_{\mathbf{p}_0}, \quad
\mathbf{J_q} \triangleq \frac{\partial f_\text{pf}}{\partial \mathbf{q}}\Bigg|_{\mathbf{q}_0}.
\end{equation}
Moving forward, we will drop the symbol $\approx$ and use equality for simplicity of notation, though we emphasize that the voltages resulting from the linearized model should be interpreted as approximations of the true voltages.

We assume that the grid-VVC dynamics reach steady-state more quickly than large changes in $\mathbf{p}$ and $\mathbf{q}_\text{d}$. 
Therefore, similar to~\cite{zhou2021}, we define constant $\tilde{\mathbf{v}}$ as  
\begin{equation}
\label{eq:vtil}
    \tilde{\mathbf{v}} \triangleq f_\text{pf}(\mathbf{p}_0, \mathbf{q}_0) + \mathbf{J_p}(\mathbf{p} - \mathbf{p}_0) - \mathbf{J_q}(\mathbf{q}_0 + \mathbf{q}_\text{d})\, ,
\end{equation}
which simplifies~\eqref{eq:LPF} to
\begin{equation}
\label{eq:linPF}
    \mathbf{v} = \mathbf{J_q q}_\text{g} + \tilde{\mathbf{v}}\,.
\end{equation}

Matrices $\mathbf{J_p}$ and $\mathbf{J_q}$ are computed using numerical approximation of the derivative of the power flow in $f_\text{pf}$ through a centered finite difference approach. 
That is, the $j$-th column of $\mathbf{J_p}$ is computed for all $j=1,\hdots, n$ by  
\begin{equation}
\label{eq:finite_diff}
    \mathbf{J}_{\mathbf{p},j} = \frac{f_\text{pf}(\mathbf{p}_0+\varepsilon \mathbf{e}_j, \mathbf{q}_0) - f_\text{pf}(\mathbf{p}_0-\varepsilon \mathbf{e}_j, \mathbf{q}_0)}{2\varepsilon}\,,
\end{equation}
where $\varepsilon \ll 1$ (e.g., $10^{-6}$), and $\mathbf{e}_j$ is the $j$-th standard basis vector in $\mathbb{R}^n$. 
A similar procedure is used to compute $\mathbf{J_q}$. 

Procedurally, this requires solving the power flow $4n$ times ($2n$ for $\mathbf{p}_0\pm \varepsilon \mathbf{e}_j$ and $2n$ for $\mathbf{q}_0\pm\varepsilon\mathbf{e}_j$), which could become computationally expensive depending on network size.
However, these computations can be parallelized and need not be updated frequently as they are only required for establishing the model for design, not for real-time control.

Because the linearization takes into account losses and does not require that voltages be close to 1~p.u., it is a more accurate model of voltages than LDF. 
Approximation accuracy is discussed in Sec.~\ref{subs:res_LDFLPF} with a case study.

\subsection{Volt-VAr Control Rule}

\begin{figure}[t]
    \centering
    \includegraphics[width=0.9\linewidth]{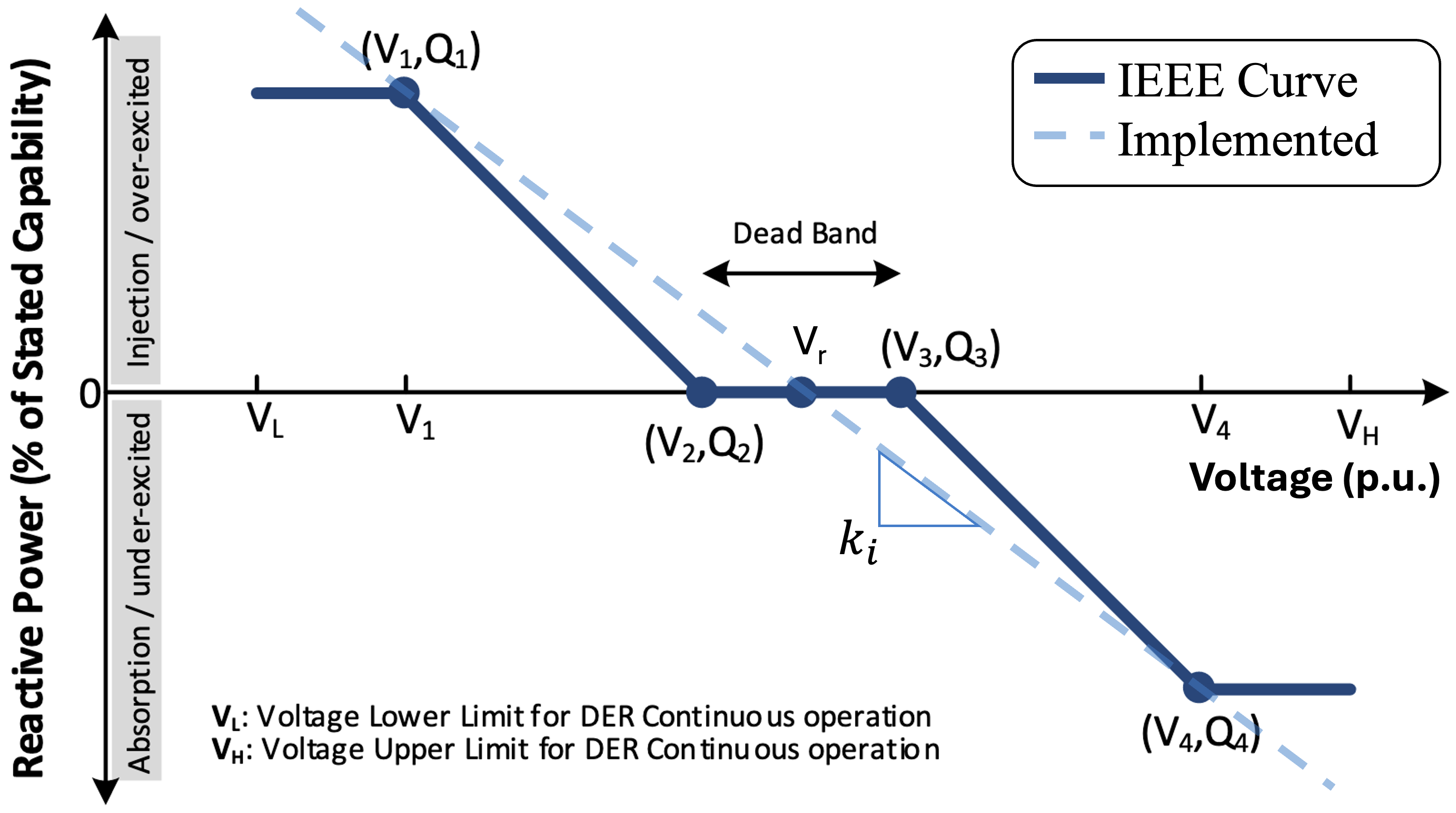}
    \caption{The IEEE prescribed standard curve with saturation limits and a deadband is shown. We choose parameters such that the curve is a straight line through the reference voltage ($V_\text{2}=V_\text{3}=V_\text{r}$ and $V_\text{L}=V_\text{1},\, V_\text{H}=V_\text{4}$). Note this is a slightly modified version of Figure H.4 from~\cite{IEEE1547}.}
    \label{fig:IEEE-VVC}
\end{figure}

The VVC curve specified by the IEEE-1547 standard~\cite{IEEE1547} is shown in Fig.~\ref{fig:IEEE-VVC}. 
The curve may include a deadband region and saturation limits, and is parameterized by the corner points $\{(V_i,Q_i)\}_{i=1}^4$ and the reference voltage $V_\text{r}$.
In this work, we consider VVC rules which do not include a deadband (i.e., $V_\text{2}=V_\text{3}=V_\text{r}$ and $Q_\text{2}=Q_\text{3}=0$), to simplify the analysis.\footnote{Inclusion of the deadband in the framework will require nonlinear control design techniques and is left as a topic of future research.}
We also assume that these nodes have sufficient reactive power capability such that saturation limits are never reached. 
Although this may be untrue in many real-world grids currently, technology such as~\cite{ecovar} is coming onto the grid that significantly increases reactive power capacity.

Under these assumptions, the resulting VVC curve at each node~$i$ is linear (as illustrated in Fig.~\ref{fig:IEEE-VVC}). 
Also, given that VVC uses sampled data, the reactive power output of VVC at time $t+1$ is dependent on voltage measured at time $t$, where $t$ is the discrete time index.
Thus, we have
\begin{equation}
    q_{\text{g},i}[t+1] = k_i(v_i[t] - v_{\text{r},i})\,,
\end{equation} 
where $v_i$ is the voltage magnitude at node~$i$, $v_{\text{r},i}$ is the reference voltage (typically 1 p.u. but can change node to node), $k_i$ is the slope of the linearized VVC curve, and $q_{\text{g},i}$ is the output reactive power. 
In this model, the size of the discrete time step $t$ is considered to be 1 second, per the lower limit of the IEEE-1547 standard open-loop response time.
 
In matrix-vector notation (for the entire network), we have
\begin{equation}
\label{eq:VVCrule}
    \mathbf{q}_\text{g}[t+1] = \mathbf{K}(\mathbf{v}[t] - \mathbf{v}_\text{r})\,,
\end{equation}
where $\mathbf{v}_\text{r} \in \mathbb{R}^n$ denotes the vector of reference voltages at each node, and $\mathbf{K}\in\mathbb{R}^{n\times n}$ is a diagonal matrix with $k_i$ on the $i$-th element of the main diagonal.
That is, $\mathbf{K}= \operatorname{diag}(\mathbf{k})$, where $\mathbf{k}=[k_1,k_2,\dots,k_n]^\top$.
For nodes that do not have an inverter, $k_i=0$.
The dimension of~\eqref{eq:VVCrule} could be reduced to include only $k_i\neq 0$, but we will need $\mathbf{q}_\text{g}$ to be in $\mathbb{R}^n$ to model effects of reactive power at non-generator nodes using~\eqref{eq:linPF}.


\subsection{Combined Grid-VVC Dynamic Model}

Next, we combine the linearized power flow equations and VVC curve to form a dynamic model of the closed-loop system, as shown in Fig.~\ref{fig:VVC_BD}. 
Substituting~\eqref{eq:VVCrule} into~\eqref{eq:linPF}, we have
\begin{align}
\label{eq:diffeq}
    \mathbf{v}[t+1] = \mathbf{J_qK}\mathbf{v}[t] - \mathbf{J_qK}\mathbf{v}_\text{r} + \tilde{\mathbf{v}}\,.
\end{align}

The equilibrium point of this discrete-time dynamical system, denoted by $\mathbf{v}^*$, is given by:
\begin{equation}
\label{eq:vstar}
    \mathbf{v}^* = (\mathbb{I} - \mathbf{J_qK})^{-1}(\tilde{\mathbf{v}} - \mathbf{J_qK}\mathbf{v}_\text{r})\,,
\end{equation}
where $\mathbb{I}$ is the identity matrix. Note that the matrix $\mathbb{I} - \mathbf{J_qK}$ is invertible when no eigenvalue of $\mathbf{J_qK}$ is $1$.
In Sec.~\ref{sec:stab_crit}, we will impose a constraint on the values of $\mathbf{K}$ that restricts all eigenvalues of $\mathbf{J_qK}$ to be strictly inside of the unit disk. 
Therefore, an equilibrium voltage solution will always exist.
The steady-state voltage equation~\eqref{eq:vstar} will be used to determine the optimal matrix $\mathbf{K}$ (and thus, the optimal VVC rules), but first we assess the stability of the system in~\eqref{eq:diffeq}.

\section{Stability Criteria}
\label{sec:stab_crit}

The discrete-time linear system~\eqref{eq:diffeq} may become unstable if VVC slopes $\mathbf{k}$ are chosen poorly. 
Thus, our goal is to select $\mathbf{k}$ such that voltages in~\eqref{eq:diffeq} are stable in time.

The standard stability criterion for a discrete-time linear system $x[t+1]=\mathbf{A}x[t]$ is given by the spectral radius $\rho(\mathbf{A})<1$. Although this is discussed in~\cite{jahangiri} with reference to VVC, they do not apply it in a design framework or use it on an actual linear model. Taking their work a step farther, we  apply it to our linear difference equation to say~\eqref{eq:diffeq} is asymptotically stable (A.S.) if this constraint is satisfied: 
\begin{equation}
\label{eq:specrad}
    \rho(\mathbf{J_qK}) < 1 \iff \text{A.S. of~\eqref{eq:diffeq}}\,.
\end{equation}

This is the least conservative stability constraint possible for a system of this type and it may be non-convex with respect to $\mathbf{k}$ depending on the structure of $\mathbf{J_q}$. 
Other work has taken a convex restriction of this constraint in order to avoid computational complexity and guarantee optimality within their reduced convex set of $\mathbf{k}$. 
In~\cite{zhou2021}, an LDF grid model is leveraged with a matrix 2-norm constraint: 
\begin{equation}
\label{eq:2norm}
    \|\mathbf{XK}\|_2 < 1 \implies \text{A.S. of~\eqref{eq:diffeq}}\,,
\end{equation}
where $\mathbf{X}$ is similar to $\mathbf{J_q}$ but calculated only from network impedances without considering losses or operating point. 
Eq.~\eqref{eq:2norm} is further restricted in~\cite{murzakhanov2024} using H\"older's inequality:
\begin{equation}
\label{eq:1infnorm}
    \|\mathbf{XK}\|_\infty < 1 \text{ and } \|\mathbf{XK}\|_1 < 1 \implies \|\mathbf{XK}\|_2 < 1\,.
\end{equation}
This constraint is unnecessarily conservative, as any induced matrix norm serves as a restriction of~\eqref{eq:specrad}, but we study it as a linear worst-case bound from the literature.

Both~\eqref{eq:2norm} and~\eqref{eq:1infnorm} avoid computational complexity at the cost of significantly restricting the set of feasible $\mathbf{k}$ values. 
We improve on these constraints by using the spectral radius combined with $\mathbf{J_q}$ from our LPF model (as opposed to $\mathbf{X}$ from the LDF model).

\begin{figure}
    \centering
    \includegraphics[width=0.9\linewidth]{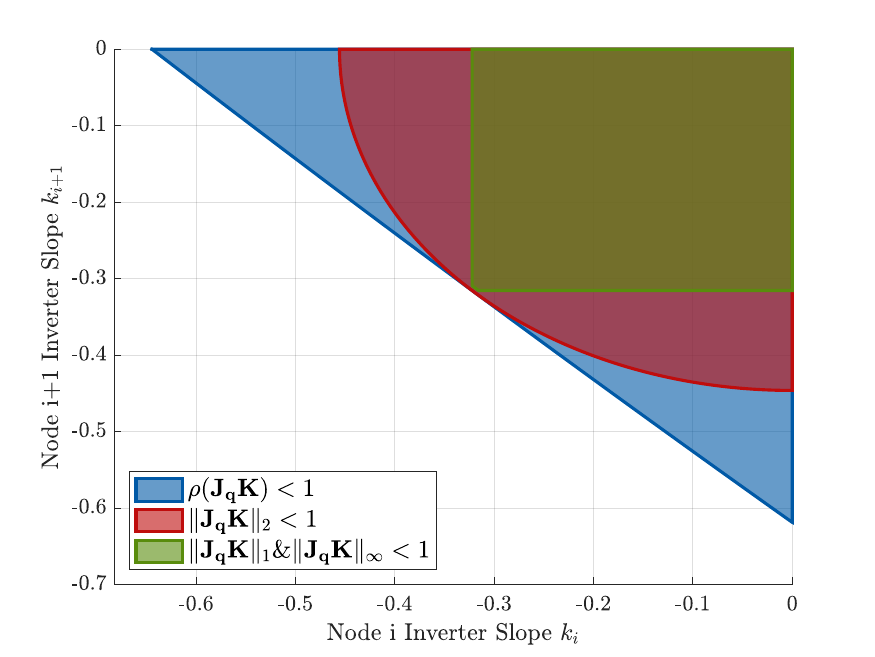}
    \caption{Comparison of stable regions of VVC slope values ($k$) in a 2-node section of a real network using an LPF model where $\mathbf{J_q}$ = [1.5504, 1.5504; 1.5505, 1.6144];. Note that the spectral radius set is convex here, although this is not guaranteed depending on the structure of $\mathbf{J_q}$.}
    \label{fig:stability_regions}
\end{figure}

A visual representation of these stability criteria is shown in Fig.~\ref{fig:stability_regions} on a 2-node section of a real feeder (see Sec.~\ref{sec:Network}). 
If both nodes have inverters with non-zero values in $\mathbf{k}$, the shaded regions represent the feasible set of asymptotically stable $\mathbf{k}$ values. 
We assume a network with inductive lines, as is common in distribution grids (and the case in this example). 
Therefore, we restrict 
$\mathbf{k} \leq 0$, 
as this will produce a negative feedback $\mathbf{q}$ response that corrects voltage towards nominal.

To the best of the author's knowledge, the spectral radius constraint has been largely avoided thus far in the literature for designing VVC slopes. 
This constraint can be non-convex depending on the structure of $\mathbf{J_q}$ which makes it more challenging to optimize over. 
Note that $\mathbf{J_q}$ is not necessarily symmetric, unlike $\mathbf{X}$ from the LDF model which is guaranteed positive definite~\cite{zhou2021}.

Intuitively, a larger $k_i$ represents greater control authority and participation in voltage regulation. To choose the best values of $k_i$, we use optimization, which is described next.

\section{Optimal VVC Design}
\label{sec:opt_design}

Now that we have established conditions on $\mathbf{k}$ values that result in stable grid-VVC dynamics, we next address how to choose the optimal $\mathbf{k}$ values from the stable set. 

In the literature, a number of network objectives have been considered in VVC design including network losses~\cite{bolognani2015} and substation power output~\cite{dakotaThesis}. 
In this work, we study voltage regulation, similar to~\cite{zhou2021}, which is critical for network reliability and flexibility. 
More specifically, we are interested in minimizing the deviation in steady-state voltage magnitudes from a given reference, $\mathbf{v}_\text{r}$.
Thus, we define the following objective function:
\begin{equation}
\label{eq:optvstar_beta}
    \min_\mathbf{k} \|\mathbf{v}^* - \mathbf{v}_\text{r}\|_2^2 + \frac{\beta}{n_\text{g}}\|\mathbf{k}\|_2^2\,, 
\end{equation}
where the steady-state voltages $\mathbf{v}^*$ are given by~\eqref{eq:vstar} and $\beta$ is a scalar cost-coefficient, normalized by $n_g$ to ensure scalability with the number of generators.
The 2-norm of voltage deviation was selected to spread the reliability and flexibility benefits of inverter VVCs across the entire network.
The regularization term limits control action by penalizing large values of $\mathbf{k}$, similar to ideas from LQR theory. 
This regularization also helps with flatness of the objective function. 

Three constraints for stability of the system were presented in Sec.~\ref{sec:stab_crit}. 
We choose to focus on the spectral radius constraint~\eqref{eq:specrad}, the least conservative, but most computationally expensive.
As discussed in Sec.~\ref{sec:stab_crit}, we also constrain each inverter slope $k_i$ to be non-positive. 
This ensures IEEE-1547 standard compliance. 
Combining these constraints with the objective function~\eqref{eq:optvstar_beta}, we obtain the following non-convex\footnote{This non-convex objective function with a possibly non-convex constraint implies there is no guarantee of global optimality.
However, we found that under a broad range of initial conditions, the solution value of~\eqref{eq:full_opt} is always lower than under constraints in~\eqref{eq:2norm}. 
Performance, as indicated by the optimal value, is more important than global optimality.} optimal VVC design formulation:
\begin{subequations}
\begin{align}
    \min_\mathbf{k} ~~&\|\mathbf{v}^* - \mathbf{v}_\text{r}\|_2^2 + \frac{\beta}{n_\text{g}}\|\mathbf{k}\|_2^2\,, \label{eq:objfn_infull}\\
    \text{s.t.}~~& \rho(\mathbf{J_qK}) \leq 1-\epsilon\,, \label{eq:stab_constraint}\\ 
    & \mathbf{v}^* = (\mathbb{I} - \mathbf{J_qK})^{-1}(\tilde{\mathbf{v}} - \mathbf{J_qK}\mathbf{v}_\text{r})\,, \label{eq:vstar_in_opt}\\
    & \mathbf{K} = \operatorname{diag}\big(\mathbf{k}\big)\,, \\
    & k_i \le 0\, , \quad \forall i \in \mathcal{N}_\text{g}\,, \\
    & k_{i}= 0\,, \quad \forall i \in \{1,\hdots, n\}\setminus \mathcal{N}_\text{g}\,.
\end{align}
\label{eq:full_opt}
\end{subequations}
To improve numerical stability, we replace $<1$ from~\eqref{eq:specrad} with $\le1-\epsilon$ in~\eqref{eq:stab_constraint}, where $0<\epsilon \ll 1$.

In order to implement the non-convex formulation in~\eqref{eq:full_opt}, we leverage MATLAB's {\tt fmincon} with default interior point algorithm. 
We utilize {\tt fmincon}'s flexibility and substitute~\eqref{eq:vstar_in_opt} into~\eqref{eq:objfn_infull}, forming a non-convex objective function due to matrix inverse and bilinearities.
Other solvers like IPOPT~\cite{ipopt} handle this by treating $\mathbf{v}^*$ as a second (dependent) variable and keeping~\eqref{eq:vstar_in_opt} as an explicit non-convex equality constraint, with $(\mathbb{I} - \mathbf{J_qK})$ brought to the left side. 
In addition, IPOPT is unable to directly express the spectral radius constraint, because of its lack of smoothness. 
Table~\ref{tab:constraint_compare} shows how the stability constraints were implemented for the different optimization solvers.

\begin{table}
\caption{Stability Constraint Implementation}
\begin{center}
\renewcommand{\arraystretch}{1.2} 
\begin{tabular}{|p{0.27\linewidth}|p{0.4\linewidth}|p{0.12\linewidth}|}
\hline
\textbf{Constraint} & \textbf{Implemented Form} & \textbf{Solver}$^{\mathrm{a}}$ \\
\hline
$\rho(\mathbf{J_qK}) < 1$ & $\max |\operatorname{eig}(\mathbf{J_qK})| \leq 1-\epsilon$  & fmincon\\
\hline
$\|\mathbf{J_qK}\|_2 < 1$ & $(\mathbf{J_qKx})^\top (\mathbf{J_qKx}) \leq 1-\epsilon$ and $\mathbf{x}^\top\mathbf{x} = 1$ & IPOPT\\
\hline
$\|\mathbf{J_qK}\|_\infty < 1$ and $\|\mathbf{J_qK}\|_1 < 1$ & $-\mathbf{J_qk} \leq (1-\epsilon)\mathbf{1}$ and $\operatorname{diag}(\mathbf{1}^\top\mathbf{J_q})\mathbf{k} \leq (1-\epsilon)\mathbf{1}$ & IPOPT \\
\hline
\multicolumn{3}{p{0.85\linewidth}}{$^{\mathrm{a}}$ Indicates the solver used, not a complete list of capable solvers.} 
\end{tabular}
\label{tab:constraint_compare}
\end{center}
\end{table}


\section{Test Data and Model Validation}
\label{sec:Network}

The solution of the optimization problem~\eqref{eq:full_opt} provides a set of VVC slopes that are guaranteed to be stable (under the LPF model) and are optimal with respect to steady-state voltage regulation. 
In order to evaluate the performance of the proposed VVC design framework, we leverage network, load, and generation data from a real distribution feeder.
Relevant details about the test case data are provided next.

\subsection{Network, Generation, and Load Data}

\begin{figure}
\includegraphics[width=\linewidth]{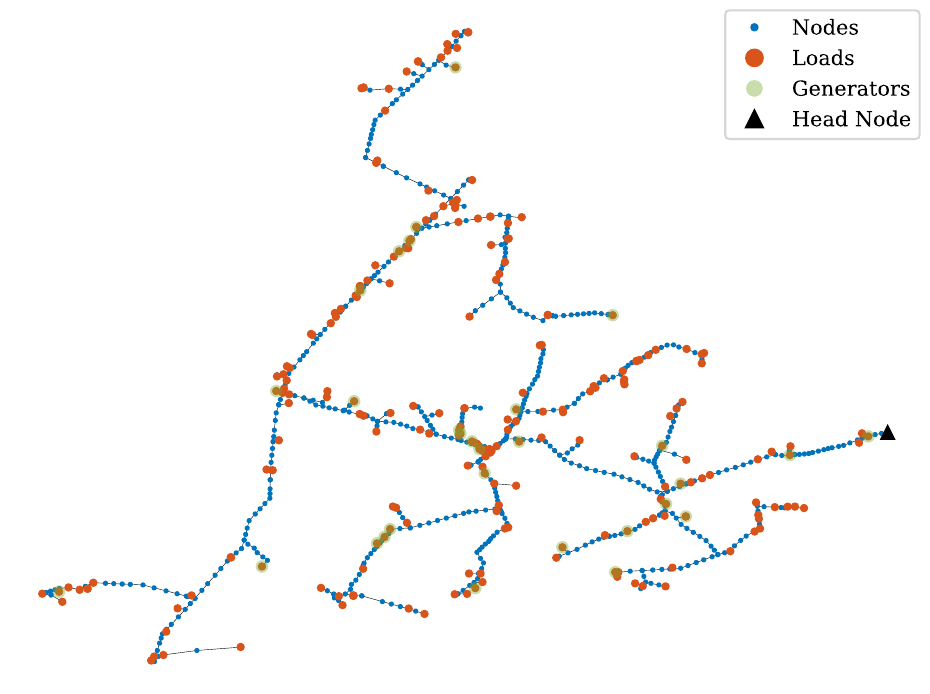}
\caption{Illustration of the single-phase section of a real feeder in Vermont on which the VVC design framework is applied and evaluated.}
\label{fig:NetworkStructure}
\end{figure}

For this study, we use a single-phase section of a real distribution feeder in Vermont, with data provided by the Vermont Electric Cooperative. 
The network contains 1072 nodes $(n=1072)$ across medium (12.47~kV) and low (208~V) voltage levels.
This feeder section contains 209 residential loads and 30 nodes with solar generation $(n_\text{g}=30)$. 
An illustration of the network can be seen in Fig.~\ref{fig:NetworkStructure}. 
While the head node of this feeder section is not the substation, it is regulated by a tap-changing voltage regulator. 
Thus, we assume that the voltage magnitude at this node is fixed at 1 per unit.
Real load and generation data from the entire year of 2024 is randomly split into a ``training dataset'' (90\%) and ``test dataset'' (10\%).
The training dataset is used to determine an average loading condition over the year, which serves as the operating point ($\mathbf{p}_0,\mathbf{q}_0$) in the LPF model~\eqref{eq:LPF}. 
The test dataset is used as sample loading conditions to which we apply our optimally designed VVC policy to assess its performance.

\subsection{Validation of Linearized Power Flow (LPF) Model }
\label{subs:res_LDFLPF}

To validate the accuracy and utility of the LPF model from Sec.~\ref{subs:LPF} for the test network, we compare it to the nonlinear AC power flow solution and the LDF model, the latter of which is commonplace throughout VVC literature. 

First, the LPF model is parameterized based on the training dataset.
Specifically, the $i$-th element of $\mathbf{p}_0$ is the average of all net active power injections at node~$i$ across all samples in the training dataset, and similarly for $\mathbf{q}_0$.
The Jacobian matrices $\mathbf{J_p}$ and $\mathbf{J_q}$ are  calculated from~\eqref{eq:finite_diff}, where the nonlinear AC  power flow solver MATPOWER~\cite{zimmerman2011} is used to evaluate $f_\text{pf}$.
We emphasize that the LPF model is constructed once (i.e., using a single operating point) and is applied to all loading conditions in the test dataset, rather than re-linearizing around a different operating point for each loading condition.

\begin{figure}
\includegraphics[width=\linewidth]{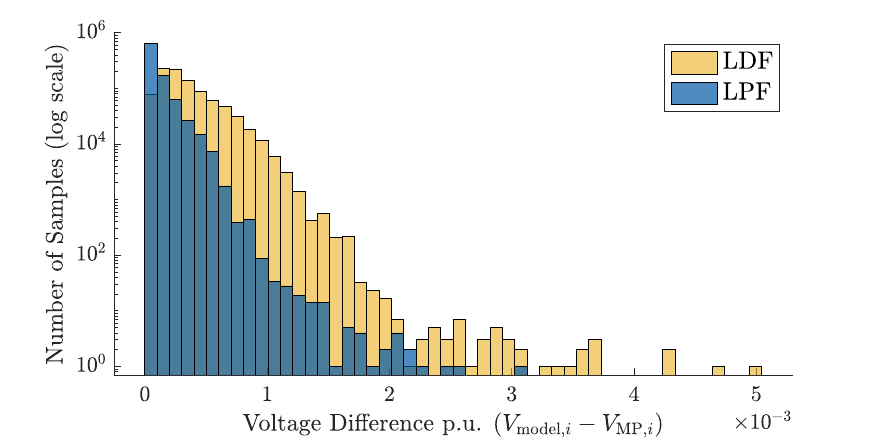}
\caption{A histogram showing the difference in nodal voltage magnitudes between the model used ($V_{\text{model},i}$ LDF or LPF) and voltage computed by MATPOWER ($V_{\text{MP}}$) across the test dataset.}
\label{fig:LDFLPF}
\end{figure}

Figure~\ref{fig:LDFLPF} compares the voltage magnitude error at each node of the LPF and LDF models ($V_\text{model}$ in figure) with respect to the nonlinear AC power flow solutions from MATPOWER ($V_\text{MP}$ in figure) across loading scenarios from the entire test dataset. 
The maximum voltage difference for LDF is $0.005$~p.u. while LPF is $0.0031$~p.u.
The results indicate that the LPF is generally more accurate than LDF for these test networks and datasets.
This improved accuracy can be attributed to LPF accounting for losses and not assuming voltage magnitudes are equal to 1 p.u. through the linearization process, both of which are reductive assumptions in the LDF model.

\section{Case Study Results}
\label{sec:Results}

Next, we demonstrate the effectiveness of the proposed VVC design framework through a number of case studies.
In particular, we compare the performance of VVC designs under different power flow models, and stability constraints in~\eqref{eq:specrad}--\eqref{eq:1infnorm}.
We also compare non-incremental VVC design to an incremental approach.

\subsection{Case Study Setup}
\label{subs:sim_meth}

The VVC curves are designed by solving the optimization problem~\eqref{eq:full_opt} under different $\beta$.
For reference voltages in the objective, we consider $\mathbf{v}_\text{r} = \mathbf{1}_n$, where $\mathbf{1}_n$ denotes a vector of ones of size $n$, and for the stability constraints, we set $\epsilon=10^{-3}$.
A specific grid loading condition $\tilde{\mathbf{v}}$ (i.e., $\mathbf{p}_\text{g}$, $\mathbf{p}_\text{d}$, $\mathbf{q}_\text{d}$) is needed for design. 
For this, we choose the largest 2-norm of voltage deviation from nominal in the training dataset, which is representative of a ``worst-case'' scenario.
Lastly, we compare the value of both terms in~\eqref{eq:full_opt} objective function under different $\beta$ and select $\beta=0.06$.

These optimal VVC curves are applied to the inverters in the network, and their performance is evaluated through numerical simulations.
Given a loading condition from the test dataset,~\eqref{eq:VVCrule} provides a new value of reactive power $\mathbf{q}_\text{g}$ to be applied at each inverter. 
We use MATPOWER to solve $f_\text{pf}(\mathbf{p},\mathbf{q})$. 
The voltage solution of $f_\text{pf}$ is fed back into VVC and the loop continues until the voltage converges to some steady state value within a tolerance of $10^{-4}$. 
By solving power flow with a nonlinear AC solver as opposed to LPF, we are closer to representing how the designed inverter slopes will perform in a real-world context.
The voltage deviation of the steady state voltage values can then be compared with different $\mathbf{k}$'s, different models, and the baseline voltage before any control action. 

\subsection{Exemplary Hours for Case Studies}

From the test dataset, we select four particular hours of the year which represent specific scenarios of interest:
\begin{itemize}
    \item Hour A: maximum total active power demand (which also corresponds to maximum voltage deviation using $\|\mathbf{v}-\mathbf{1}_n\|_2$).
    \item Hour B: maximum total active power generation.
    \item Hour C: minimum single-node voltage.
    \item Hour D: maximum single-node voltage.
\end{itemize}
These four hours are representative of the range of loading conditions in the full test dataset. 

\subsection{Comparing Models}
\begin{figure}
    \centering
    \includegraphics[width=\linewidth]{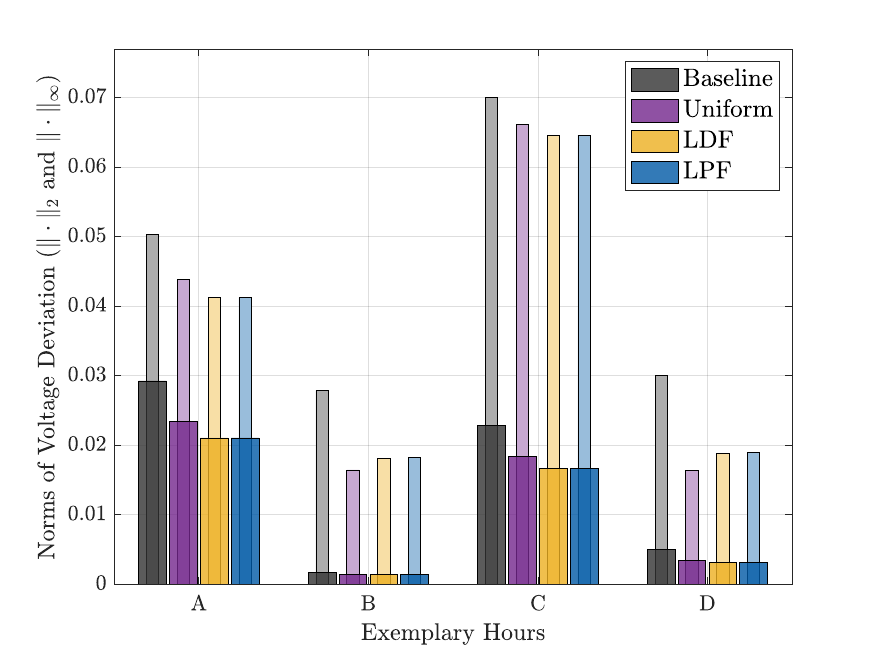}
    \caption{A comparison of the performance of VVC under different design techniques: arbitrary uniform values, optimal with LDF, and optimal with LPF. 
    There is no meaningful difference between LDF and LPF. The uniform values perform better under low voltage deviation (B, D) because the optimal VVC is over-correcting. 
    The 2-norm of voltage deviation (darker wider bar) and $\infty$-norm of voltage deviation (lighter skinnier bar) are both shown defined as $\|\mathbf{v}-\mathbf{v}_\text{r}\|$.}
    \label{fig:ModelCompare}
\end{figure}

We first compare the performance of optimal $\mathbf{k}$ values under different model choices. 
In Fig.~\ref{fig:ModelCompare}, the baseline is represented in gray. 
We also show a ``model-free'' case in purple, where all inverter slopes are set to a uniform (guaranteed stable) value. 
Voltage deviation from optimal slopes designed using the LPF and LDF models are shown in blue and yellow respectively. 

In Sec.~\ref{subs:res_LDFLPF}, we showed with Fig.~\ref{fig:LDFLPF} that LPF is generally a more accurate model of voltage than LDF. 
Despite this, the $\mathbf{k}$ values designed by the two models perform almost identically when correcting voltage deviation. 
This indicates that, under these loading conditions and this network, both LDF and LPF are suitable grid models.

We also observe that designing the $\mathbf{k}$ values via either optimization problem is more effective than setting all inverters to be arbitrarily the same value when voltage deviation is high. 
In low voltage deviation situations, the stronger control action from optimal $\mathbf{k}$ is over-correcting at nodes nearby to the inverter. 
We expect that implementation of a deadband would eliminate this effect, which is a direction of future work. 


It is worth noting that all three implementations of VVC will reduce voltage deviation, including if $\mathbf{k}$ values are set uniformly across the network. 
The control policy doesn't have to be ``optimal'' to improve the voltage profile. 

\subsection{Performance of Stability Constraints}

\begin{figure}
    \centering
    \includegraphics[width=\linewidth]{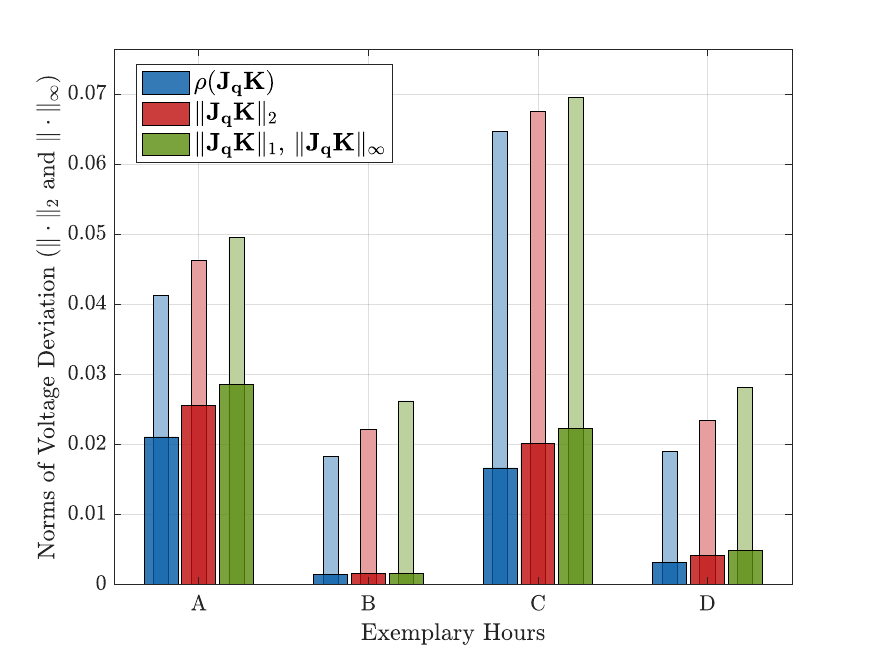}
    \caption{A performance comparison of optimal $\mathbf{k}$ values designed using different stability criteria in place of~\eqref{eq:stab_constraint}. As expected, more restrictive constraints perform worse. The 2-norm of voltage deviation (darker wider bar) and $\infty$-norm of voltage deviation (lighter skinnier bar) are both shown.}
    \label{fig:ConstraintCompare}
\end{figure}

In Fig.~\ref{fig:ConstraintCompare}, we compare the stability constraints that were discussed in Section~\ref{sec:stab_crit}. 
To generate these results, we solve the optimization in~\eqref{eq:full_opt}, replacing~\eqref{eq:stab_constraint} with different stability criteria. 
The optimal $\mathbf{k}$ values are used to simulate the control policy performance for hours A--D. 
Note that these slope values were all designed using the LPF model.

It is not surprising that the most restrictive stability criterion performs the worst. 
This trend is noticeable even under relatively low voltage deviation like in hour B. 

\begin{figure}
    \centering
    \begin{tikzpicture}
    \node[inner sep=0pt] (n1) at (0,0) {\includegraphics[width=\linewidth]{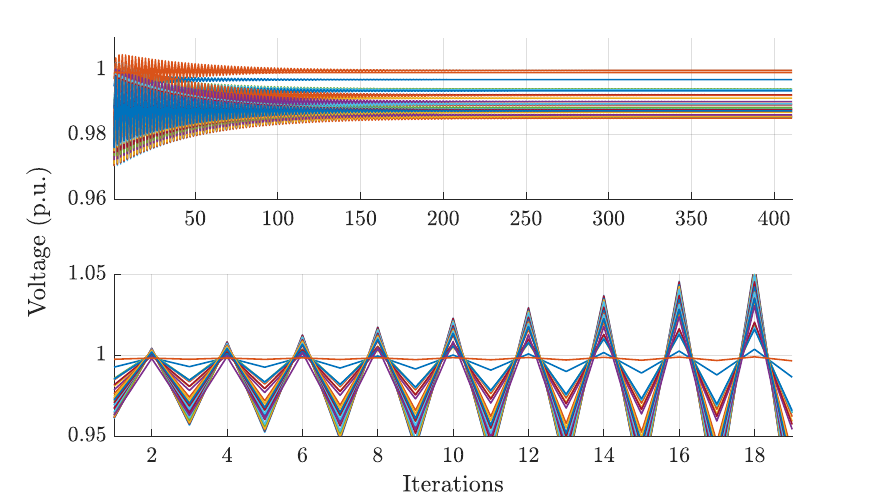}};
    \node[inner sep=0pt] (n1) at (-4.0,1.5)
    {\small (a)};
    \node[] (n2) at (-4.0,-1.25) {\small (b)};
    \end{tikzpicture}
    \caption{The voltage dynamics at generator nodes are shown, simulated by MATPOWER during hour A loading conditions; (a) Spectral radius constraint --- stable dynamics and (b) $\mathbf{k}$ increased by 10\% --- unstable dynamics. Note that faster convergence can be guaranteed by increasing $\epsilon$ in~\eqref{eq:stab_constraint}.}
    \label{fig:VoltageOscillations}
\end{figure}

In Fig.~\ref{fig:VoltageOscillations}(a), the simulated voltage dynamics for hour A at all inverter nodes are shown, using $\mathbf{k}$ designed under the spectral radius stability constraint.
To show instability, we increase $\mathbf{k}$ by 10\%, shown in Fig.~\ref{fig:VoltageOscillations}(b).
These dynamics are simulated as described in Sec.~\ref{subs:sim_meth}. 
With unstable slopes for inverters, the voltage oscillates until exceeding physical limits of the grid.
Voltage dynamics under the other two constraints are similar, with slightly faster convergence and less voltage regulation. 
They also generally have more room to increase $\mathbf{k}$ before becoming unstable. 

\subsection{Comparison with Incremental}

\begin{figure}
    \centering
    \includegraphics[width=\linewidth]{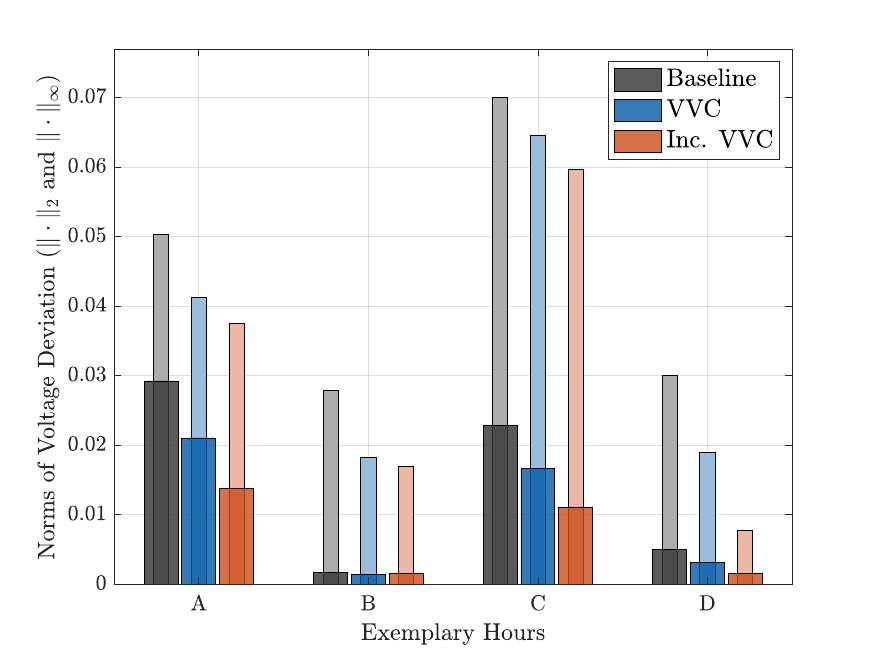}
    \caption{A comparison of 4 different hours of the year under non-incremental (VVC) and incremental (Inc. VVC) control. The 2-norm of voltage deviation (darker wider bar) and $\infty$-norm of voltage deviation (lighter skinnier bar) are both shown.}
    \label{fig:VoltDevResult}
\end{figure}

Finally, we compare our designed non-incremental solution to an incremental VVC approach. 
The authors in~\cite{eggli2021} prove that high gain values which would otherwise destabilize the grid can be stabilized using an incremental approach. 
To demonstrate this, we arbitrarily take our designed $\mathbf{k}$ values and multiply them by a factor of 25, putting them well outside of the spectral radius stability boundary. 
We implement a ``Type-B'' system from~\cite{jahangiri} with $\frac{\Delta T}{\tau}=0.05$. 
Results in~\cite{jahangiri} demonstrate that, if $\frac{\Delta T}{\tau} \leq 1$, the design framework presented here also applies to an incremental implementation.\footnote{Note that the special case of $\frac{\Delta T}{\tau}=1$ is the rule in~\eqref{eq:VVCrule}.} 
Simulation results of using this incremental approach can be seen in Fig.~\ref{fig:VoltDevResult} as ``Inc. VVC''. 

\begin{figure}
    \centering
    \includegraphics[width=\linewidth]{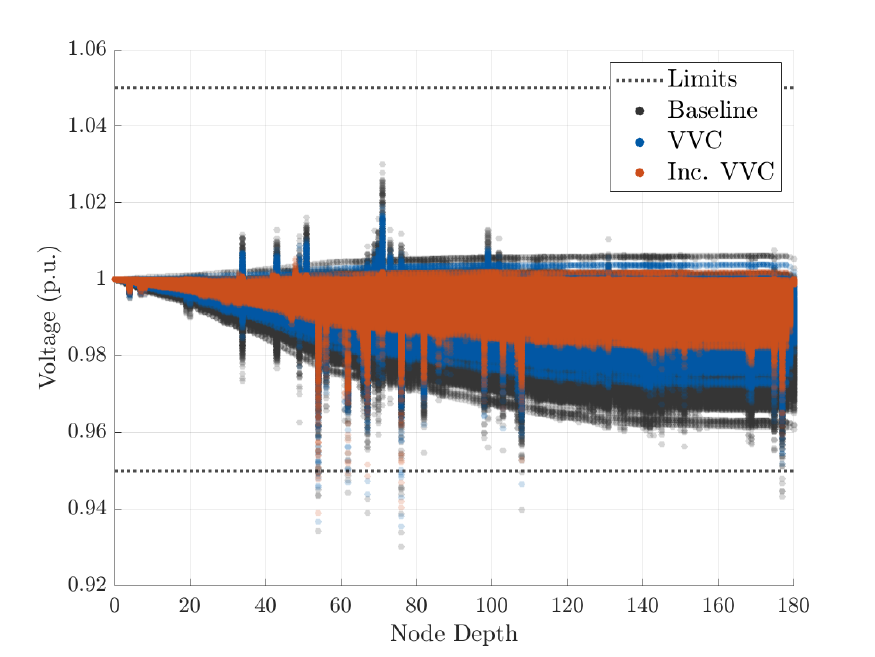}
    \caption{Voltage at all nodes over the entire test dataset, plotted under different control schemes (no control, VVC, incremental VVC). Spacing on the x-axis is based on how many nodes are between the plotted node and the head node.}
    \label{fig:VoltageCloud}
\end{figure}

An incremental approach correcting the voltage so much more than our design demonstrates that the stability constraint is a fundamental limit to voltage regulation performance. 
Fig.~\ref{fig:VoltageCloud} further illustrates the corrective power of both control policies.
This figure shows the voltage at all nodes across the entire test dataset of loading conditions with baseline, non-incremental, and incremental VVC. 
Voltage at each node is ordered on the x-axis according to how many nodes are between that node and the head node. 
We observe that, in general, incremental VVC brings voltages closer to nominal.
It is worth noting that the average number of iterations for non-incremental to converge is 389, 
while incremental converges in 9.4 iterations on average with 25 times higher slopes.

Some nodes are not being corrected sufficiently, even with the incremental VVC response, because the generators are too far away from affected nodes. 
One way to correct this is to add a generator at a particular node with high voltage deviation, re-optimize over $\mathbf{k}$ and re-simulate the grid. 
As a simple test of the effect of this, we add an inverter at the node causing significant voltage drop at depth 76. 
From Fig.~\ref{fig:VoltDevResult} during hour~C, regular VVC has a voltage deviation infinity norm of 0.065, which violates the limits set by ANSI standards (0.05). 
After adding an inverter at that node, the infinity norm of voltage deviation during that same hour drops to 0.040, which is well within the limits. 



\section{Conclusions}
\label{sec:conclusion}

In this study, we compared different stability constraints and grid models while finding an optimal design of inverter slopes for VVC, and evaluated our design framework using a real network with real loading conditions. 
We conclude that Linearized Power Flow is generally a more accurate tool for modeling voltage than LinDistFlow on a real network using real data. 
However, this increased accuracy does not necessarily translate to better performance when designing VVC rules. 
Therefore, either model is suitable for design. We also showed that a simple VVC design using uniform slopes across the network has the ability to improve voltage deviation, although not as much as optimized design.

Different stability constraints were studied and we showed that although a spectral radius constraint on inverter slopes is more challenging to compute, it produces  a more optimal result when used for VVC design. 
We also showed that an incremental approach which allows for larger slopes without stability issues often results in less voltage deviation. 

Future work will focus on optimizing design of incremental VVC, including with deadband and saturation limits. 
The question of controllability is also of interest, given that this system has a limited number of nodes that have active control. 
It would be valuable to understand where the optimal location would be to add reactive power capability.
Lastly, we aim to extend these results to 3-phase networks.

\bibliographystyle{ieeetr}
\bibliography{Bib07}

\end{document}